\documentstyle[12pt,twoside,fleqn,espcrc1,epsf]{article}

\newcommand{\AmS}{{\protect\the\textfont2
  A\kern-.1667em\lower.5ex\hbox{M}\kern-.125emS}}

\title{Scattering states in the four nucleons system}

\author{F. Ciesielski, J. Carbonell, C.Gignoux
\address{Institut des Sciences Nucl\'eaires, 53 Av. des martyrs, 38026 Grenoble, France } }

\begin{document}
\maketitle

\begin{abstract}
The Faddeev-Yakubowski equations have been solved in configuration space for the four nucleons system.
Results for bound and scattering states in the isospin and S-wave approximation for different
(T,S) channels are presented.
The n-t elastic cross section has been also calculated with realistic interactions.
A special interest was devoted to the description of the observed resonant structure at $T_{lab}\approx 3$ MeV
which is well reproduced only by the simple MT I-III potential.
\end{abstract}

\section{INTRODUCTION}

We present in this contribution the results obtained by solving the Faddeev-Yakubowski (FY) equations in configuration
space for the 4N system with special interest in their scattering states.
The A=4 chart presents a very rich continuum and constitutes a challenge for the theoretical models.
We have limited our study to the energy region below the first inelastic channel.
All the calculations have been carried out in the isospin approximation, i.e neglecting Coulomb forces.
Next section  will be devoted to a short presentation of the FY formalism.
In Section 3 we will discuss the results: first those obtained in the S-wave approximation
for the different (T,S) channels
and secondly those concerning the n-t scattering with realistic interactions.

\section{FORMALISM}

The FY equations for 4 interacting particles are obtained by first splitting the total wavefunction $\Psi$
in the usual Faddeev amplitudes, $\Psi_{ij}$, associated with each of the interacting pair.
Each of these amplitudes is in its turn splitted
in 3 parts, the FY amplitudes, corresponding to the different asymptotics of the remaining two non interacting particles.
The Schr\"odinger equation is equivalent to the set of 18 coupled equations 
for which it is possible to define appropriate boundary conditions ensuring the unicity of the solution.

In the case of identical particles, the 18 FY amplitudes can be obtained by the action of the
permutation operators $P_{ij}$ on two of them, denoted $K$ and $H$. 
The $K-H$ amplitudes satisfy the following equations:
\begin{eqnarray}
(E-H_0-V)K &=&V\left[ (P_{23}+P_{13})\;(\varepsilon + P_{34})\; K + \varepsilon(P_{23}+P_{13})\; H\right]\label{FYE1}\\
(E-H_0-V)H &=&V\left[ (P_{13}P_{24}+P_{14}P_{23}) \;K +    P_{13}P_{24} \; H\right]  \label{FYE2}
\end{eqnarray}
in which $\varepsilon=\pm1$ depending on whether we are dealing with bosons or fermions.
Each amplitude $F=K,H$ is considered as function of its own set of Jacobi coordinates
$\vec{x},\vec{y},\vec{z}$
and expanded in angular momentum variables for each coordinate according to
\begin{equation}\label{KPW}
<\vec{x}\vec{y}\vec{z}|F>=
\sum_{\alpha} \; {F_{\alpha}(xyz)\over xyz} \; Y_{\alpha} (\hat{x},\hat{y},\hat{z}) 
\end{equation}
where $Y_{\alpha}$ are generalized tripolar harmonics containing spin, isospin and angular momentum variables
and the functions $F_{\alpha}$, called FY components, are the unknowns.
The label $\alpha$ holds for the set of intermediate quantum numbers defined in a given coupling scheme
and includes the specification for the type of amplitudes K or H. We have used the following couplings:
\begin{eqnarray*}
\mbox{K amplitudes}  & &\left\{ \left[ (t_1 t_2)_{\tau_x} t_3 \right]_{T_3} t_4 \right\}_T \otimes
\left\{ \left[ \left( l_x (s_1 s_2)_{\sigma_x} \right)_{j_x} (l_y s_3)_{j_y}
\right]_{J_3} (l_z s_4)_{j_z} \right\}_J     \cr
\mbox{H amplitudes}  & & \left[ (t_1 t_2)_{\tau_x} (t_3 t_4)_{\tau_y} \right]_T \otimes
\left\{ \left[ \left( l_x (s_1 s_2)_{\sigma_x} \right)_{j_x} 
\left( l_y (s_3 s_4)_{\sigma_y} \right)_{j_y} \right]_{j_{xy}} l_z \right\}_J 
\end{eqnarray*}
where  $t_i$ and $s_i$ are the isospin and spin of the individual particles 
and $(T,J)$ are respectively the total isospin and angular momentum of the four-body system.
Each component $F_{\alpha}$ is thus labelled  by a set of 12 quantum numbers
to which the symmetry properties of the wavefunction impose the additional constraints:
$(-1)^{\sigma_x+\tau_x+l_x}=\varepsilon$ for K  
and $(-1)^{\sigma_x+\tau_x+l_x}=(-1)^{\sigma_y+\tau_y+l_y}=\varepsilon$ for H.

The boundary conditions are implemented by imposing at large enough values 
of $z$ the asymptotic behaviour of the solutions.
Indeed, at large values of $z$ and for energies below the first inelastic threshold, the solution of equation (\ref{FYE1}) 
factorizes into a bound state solution of the 3N Faddeev equations and a plane wave in the $z$ direction
whereas the solution of (\ref{FYE2}) vanishes.
The scattering observables are directly extracted from the logarithmic derivative of the 
K amplitude in the asymptotic region.
The asymptotic 3N state is calculated with the same numerical scheme than the one used to solve the 4-body problem. 
In this case the factorization property, valid only in cartesian coordinates, is 
an exact numerical property and provides a strong stability test  \cite {These}.

The numerical methods used are based on the Hermite spline expansion, orthogonal collocation
and iterative procedures for solving the linear system.
An important step in its solution is done by means of the tensor trick \cite{SSK 92}.

\section{RESULTS}

\subsection{S-wave approximation}

We first discuss the results obtained in the S-wave approximation.
By this we mean i) an interaction which acts only on S-wave and 
ii) that all the angular momenta variables in expansion (\ref{KPW}) are zero.
The interaction considered is MT I-III with parameters given in \cite{LA 82}, 
slightly different from those of its original version \cite{MT 69}.
The scattering length and effective range in different 
(T,S) channels are given in  Table~\ref{tab1} and compared with the existing calculations.
\begin{table}[hbt]
\begin{minipage}[h]{79mm}
\caption{Low energy N-NNN parameters}\label{tab1}
\begin{tabular}{cc r rr} \hline
T & S & a     &  $r_0$ &  a \cite{YF 95}	     \\  \hline
0 & 0 & 14.75 &  6.75  &  $^*$14.7 	   \\
0 & 1 &  3.25 &  1.82  &  $^*$2.8 	   \\
1 & 0 &  4.13 &  2.01  &	  4.0		   \\
1 & 1 &  3.73 &  1.87  &	  3.6		   \\
\hline\end{tabular}
\end{minipage}
\hspace{\fill}
\begin{minipage}[h]{79mm}
\caption{$^4$He binding energy and r.m.s. radius}\label{tab2}
\begin{tabular}{crrrr} \hline 
       &  $^4He$ &  \cite{SSK 92} & \cite{KG 92} & $^4He^*$ 	\\  \hline
 B     & 30.30   &  30.31         &  30.29       &  0.27    \\
r.m.s. &  1.44   &   -	          &   -	         &  4.99      \\
\hline\end{tabular}
\end{minipage}
\end{table}
Our scattering length values are in agreement
with \cite{YF 95} except in the S=1 T=0 channel.
For T=1 they are also close to those obtained in \cite{Tjon 76} although with the original potential parameters.

\subsection{n-t scattering with realistic interactions}

A step behind the S-wave approximation has been achieved by considering the n-t scattering
with realistic NN potentials.
This reaction is a pure T=1 channel, free from the difficulties related to the Coulomb interaction
and for which accurate low energy  scattering data exist \cite{PBS 80}.
The extrapolated zero energy cross section is $\sigma(0)=170.2\pm 9.7$  fm$^2$ but
the spin-dependent scattering length values are still controverted (see Table~\ref{tab3}).

We have calculated the scattering length of the singlet $0^+$ and triplet $1^+$ states with
the AV14 \cite{AV14 84} and Nijmegen II \cite{NIJ 93} potentials.
The interaction is limited to the $^1S_0,^3S_1,^3D_1$ waves and the FY expansion includes all the
components with $l_z=0,1$ and $l_y=0,1,2$.
The results are displayed in Table~\ref{tab4} together with the $^4$He binding energy ($B_4$) and r.m.s. radius.
They are close to those presented in this Conference by the Pisa group \cite{Viviani}
and our $B_4$ value for AV14 shows a relative agreement of $10^{-4}$ with \cite{KG 92}. 

\begin{table}[t]
\begin{minipage}[t]{64mm}
\caption{Experimental n-t scattering lengths (in fm.).}\label{tab3}
$\begin{array}{cc}\hline
   a_0       &  a_1	    	       \\ \hline
3.91\pm0.12  & 3.6\pm0.1     \;  \cite{SL1}	 \\ 
4.98\pm0.29  & 3.13\pm0.11   \;  \cite{SL2}    \\ 
2.10\pm0.31  & 4.05\pm0.09   \;  \cite{SL2}    \\ 
4.453\pm0.10 & 3.325\pm0.016 \;  \cite{SL3}  \\\hline
 \end{array}$
\end{minipage}
\hspace{-0.0cm}
\begin{minipage}[t]{89mm}
\caption{Calculated n-t scattering lengths, zero-energy cross section and $^4$He binding energy ($B_4$) and
 r.m.s. radius (units are fm and MeV)}\label{tab4}
$\begin{array}{l rr c rr}\hline
                &  a_{0^{+}} & a_{1^{+}} & \sigma(0)      & B_4   & r.m.s.    \\ \hline
\hbox{AV14}     &   4.31     & 3.79      &  193.7         & 23.34 & 1.56  \\  
\hbox{Nijm II}  &   4.31     & 3.76      &  191.6         & 23.39 & 1.54 \\  
\hbox{MT I-III} &   4.10     & 3.63      &  177.0         &       &	  \\  
\hbox{exp}      &            &           &  170.2\pm 9.7  &       &	  \\ \hline
\end{array}$
\end{minipage}
\end{table}

The MT I-III results presented in the previous section
give a sligthly overestimated value for the zero energy cross section $\sigma(0)=184.6$  fm$^2$.
However the inclusion of higher partial waves in the FY expansion reduces this value to $\sigma(0)=177.0$  fm$^2$
in closer agreement with experiment. The corresponding scattering lengths are then $a_0=4.10$, $a_1=3.63$ fm. 

The conclusion is that the considered realistic interactions on one hand give results very close to each other
and on another hand fail in describing the zero energy n-t scattering as they failed in the 3N case.
These problems are strongly correlated and can be solved by including a three-nucleon interaction (TNI).
We have considered a simple phenomenological TNI model 
\[ W(\rho)=W_r {e^{-2\mu\rho}\over\rho}-W_a {e^{-\mu\rho}\over\rho} \qquad \rho=\sqrt{x^2+y^2} \]
and get the overall agreement $B_3=8.48$ MeV, $B_4=28.56$ MeV, $a_0=4.0$ fm, $a_1=3.6$ fm and $\sigma(0)=172.4$ fm$^2$
with the parameters $W_r=500$ MeV, $W_a=174$ MeV and $\mu=2.0$ fm$^{-1}$.
This set of values is far from being unique and the only serious constraint is the existence of a repulsive short 
range part.

The calculations have been completed up to the first 3-body break-up threshold by including 
the first negative parity states $J^{\pi}=0^-,1^-,2^-$ corresponding to a n-t relative P-waves.
This region ($T_{lab}\approx 3$ MeV) is of particular interest due to the observed structure
which has been interpreted as a series of P-waves resonances \cite{Hale 92}.

\begin{figure}[htb]
\begin{minipage}[t]{79mm}
\epsfxsize=7.9cm\epsfysize=8.0cm\mbox{\epsffile{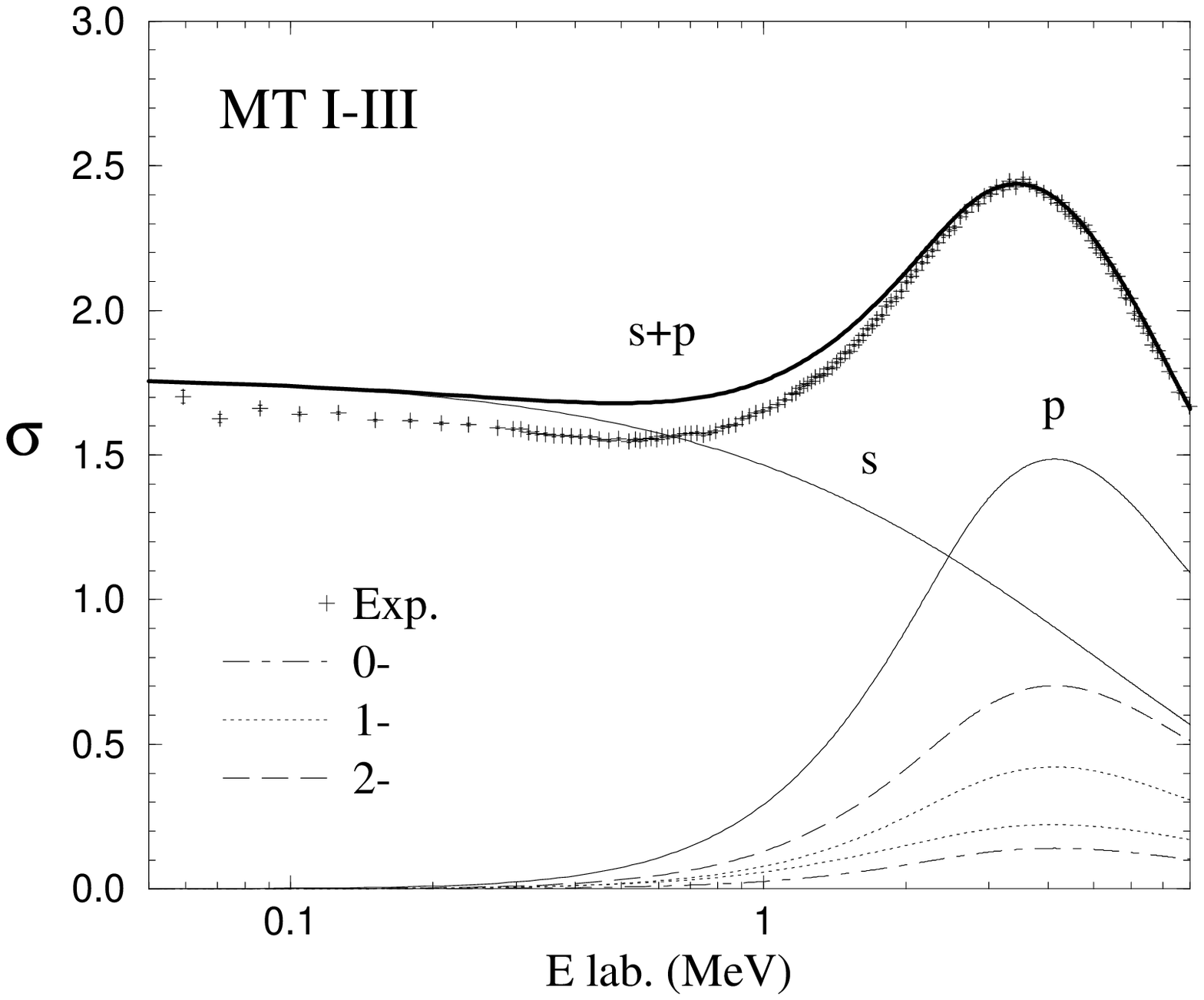}}
\caption{n-t cross section (MT~I-III).}\label{nt1}
\end{minipage}
\hspace{\fill}
\begin{minipage}[t]{79mm}
\epsfxsize=7.9cm\epsfysize=8.0cm\mbox{\epsffile{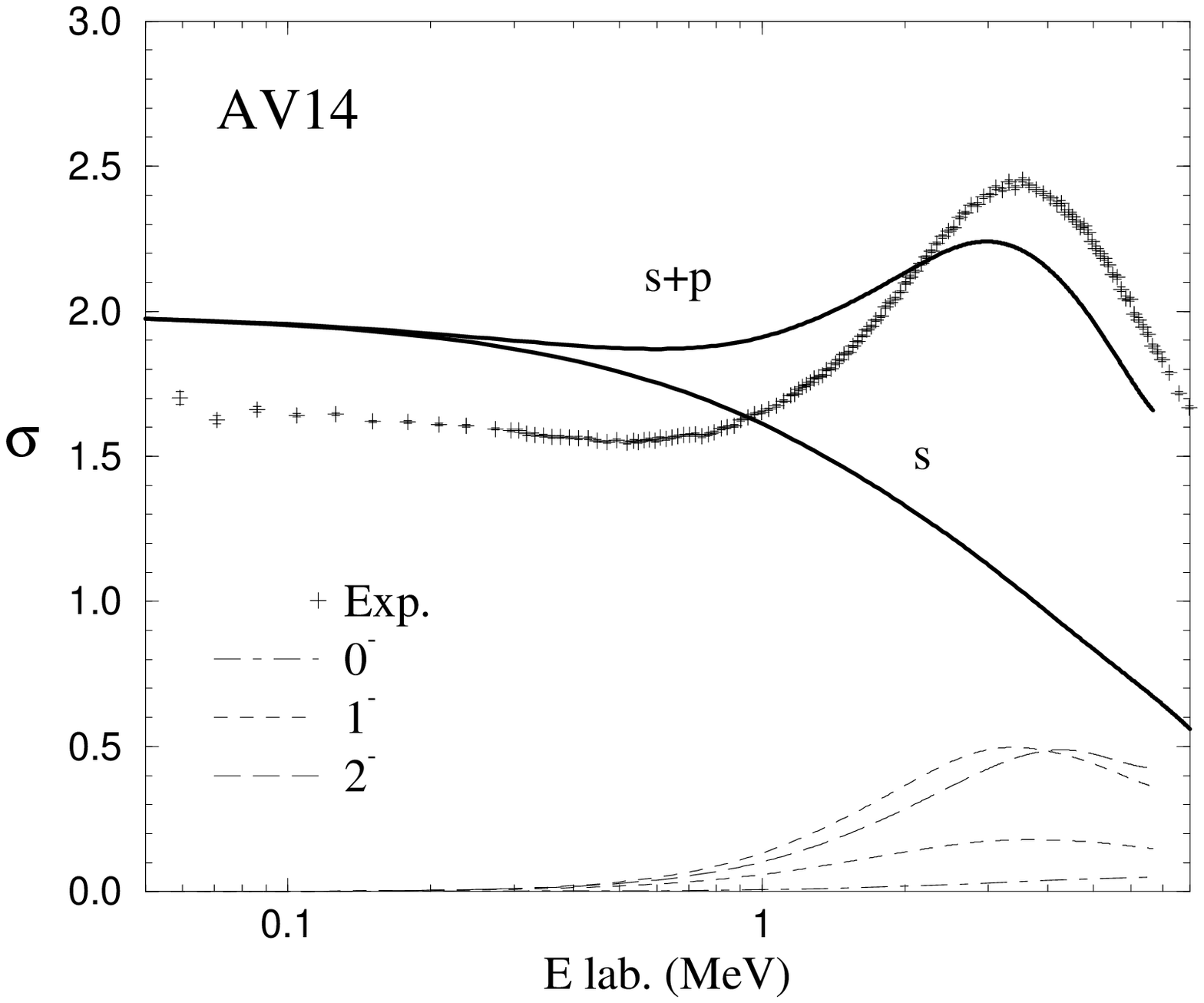}}
\caption{n-t cross section (AV14).}\label{nt2}
\end{minipage}
\end{figure}

Let us first consider the results obtained with MT I-III, displayed in Figure~\ref{nt1}.
This model conserves separately L and S and consequently
the $J^{\pi}=0^-,1^-,2^-$ states coming from S=1 are degenerated.
The corresponding cross sections differ only by statistical factors.
The remaining  $J^{\pi}=1^-$ state comes from an $(S=0,L=1)$ coupling.
Despite the simplicity of this model the agreement with experimental data,
specially in the resonance region, is very good. Only the zero energy region is a little overestimated.
The n-t P-waves resonances are thus generated by S-wave NN interaction alone.
The effective 1+3 potential is essentially due to the exchange mechanism between the four nucleons.
We notice however that a first attempt in describing this cross section with $l_z=1$ in the K components but
keeping zero all the remaining angular momenta in (\ref{KPW}) failed.

The results obtained with the AV14 are plotted in Figure~\ref{nt2}.
They are not able to reproduce the experimental data neither in the S-wave nor in the resonance region.
Correcting the zero energy with our TNI model results in lowering
the value of the cross sections and the disagreement becomes worse.
When compared to the MT I-III results one sees that, 
due to non central terms in the NN potential, the different $J^{\pi}$ contributions are shifted.
As a consequence the coherent enhancement is weakened and the resonance pick broadened.
On another hand the $2^{-}$ partial cross section, which should provide the dominant contribution, is noticeably smaller.
Trying to account for this disagreement we have included the
$^1P_0,^1P_1,^3P_1$ waves in $V_{NN}$, closing  the $J=1$ shell.
The results, displayed in Table~\ref{tab5}, show that the corrections go in the bad direction.
\begin{table}
\caption{P-wave n-t phase shifts at E=3 MeV (degrees).}\label{tab5}
$
\begin{array}{llcccc}\hline
                &    V_{NN} \quad\mbox{partial waves}  &  0^-  & 1^-  & 1^-  & 2^-	\\
\hbox{AV14}     & ^1S_0,^3S_1,^3D_1                    &  21.2 & 26.6 & 47.8 & 35.4	\\
                & ^1S_0,^3S_1,^3D_1+^1P_0,^1P_1,^3P_1  &  24.1 & 20.0 & 37.5 & 32.7   \\
\hbox{MT I-III} & ^1S_0,^3S_1                          &  44.1 & 30.3 & 44.1 & 44.1	\\
\hline\end{array}
$
\end{table}
On the contrary if the $^3S_1-^3D_1$ coupling is decreased and $V_{^3S_1}$ enhanced, keeping constant the triton binding
energy, the experimental values are reproduced.

The results obtained with MT I-III show the difficulty
to understand these resonances in terms of the NN interaction alone.
Several reasons can be put forward to explain the apparent difficulty
for realistic interactions in describing the n-t scattering before having to conclude about its definite failure.
Complementary calculations would be suitable
with increasing number of FY components, by adding
higher partial waves in the potential, by including realistic TNI, etc. 
They are however in the limit of the capacities of the best existing computers.

On another hand it is intriguing that very simple interactions work so well.
Is it only a chance that the conjugated actions of very sophisticated forces
\(V_{2N}+V_{3N}+\ldots \) result in a very simple two-body effective potential ?.
Or should that rather be an indication that simple effective potentials 
must be introduced from the very beginning in these low energy few-nucleons physics ?.

\bigskip
{\bf Acknowledgements.} The numerical part of this work has been done with the T3E
                   of the CGCV from C.E.A. and IDRIS from C.N.R.S.

\end{document}